\documentclass{eas}
\usepackage{graphicx}
\def\farcs{\hbox{$.\!\!^{\prime\prime}$}}
\def\arcmin{\hbox{$^\prime$}}
\def\arcsec{\hbox{$^{\prime\prime}$}}
\def\farcm{\hbox{$.\mkern-4mu^\prime$}}
\begin{document}

\title{An X-ray View of Radio Sources} 
\author{D. A. Schwartz}\address{Harvard/Smithsonian Center for Astrophysics, Cambridge, MA 02138, USA}
%
%
\begin{abstract}
  We review recent examples where the synergy between radio and X-ray
  observations has led to substantial progress in understanding
  astronomical systems. The sub-arcsecond imaging capabilities of the
  \emph{ Chandra} X-ray observatory provides a 100-fold improvement
  for comparing X-ray and radio structures. We specifically discuss
  examples which provide insight into the outflow of material and
  energy from pulsars and supernovae, the centers of clusters of
  galaxies, and the nuclei of quasars.
\end{abstract}

\maketitle
\section{Introduction \label{intro}}
 
In the X-ray band, as in the radio region, one studies virtually every
type of astronomical system.  Both channels are good indicators of
non-thermal activity, and of the beginnings and ends of the lifecycle
of stars and massive black holes.  In the early days of X-ray
astronomy, when locations of sources were no more precise than the
order of a square degree, the coincidence of a strong radio source in
the region served as an argument for the X-ray identification, for
example, M87 (\cite {Bradt67}) and 3C 273 ( \cite{Bowyer70}).

As a theme for this review, I will concentrate on systems where the
X-ray and radio observations are telling us about the outflow of
material and energy.  This is in contrast to what might be called
``classical'' X-ray astronomy, which studied accretion processes.
Infall of gas not only explained the energy source, but also led
directly to the calculations of X-ray emission from disks in galactic
neutron star, black hole, and white dwarf binaries. Accretion onto
supermassive black holes was inferred to power quasars and Active
Galactic Nuclei (AGN). In these cases the X-ray and non-thermal
optical continuum are most closely related.  If accretion disks were
the only setting for X-ray emission, there might be few X-ray and
radio connections, as the peak radiation frequency corresponds to the
temperature in the innermost disk, and therefore does not span the
very broad range from $\approx 10^{9}$ Hz radio emission to $\approx
10^{18}$ Hz X-ray emission.

\begin{figure}[t]
\includegraphics*[
width=.96\textwidth]{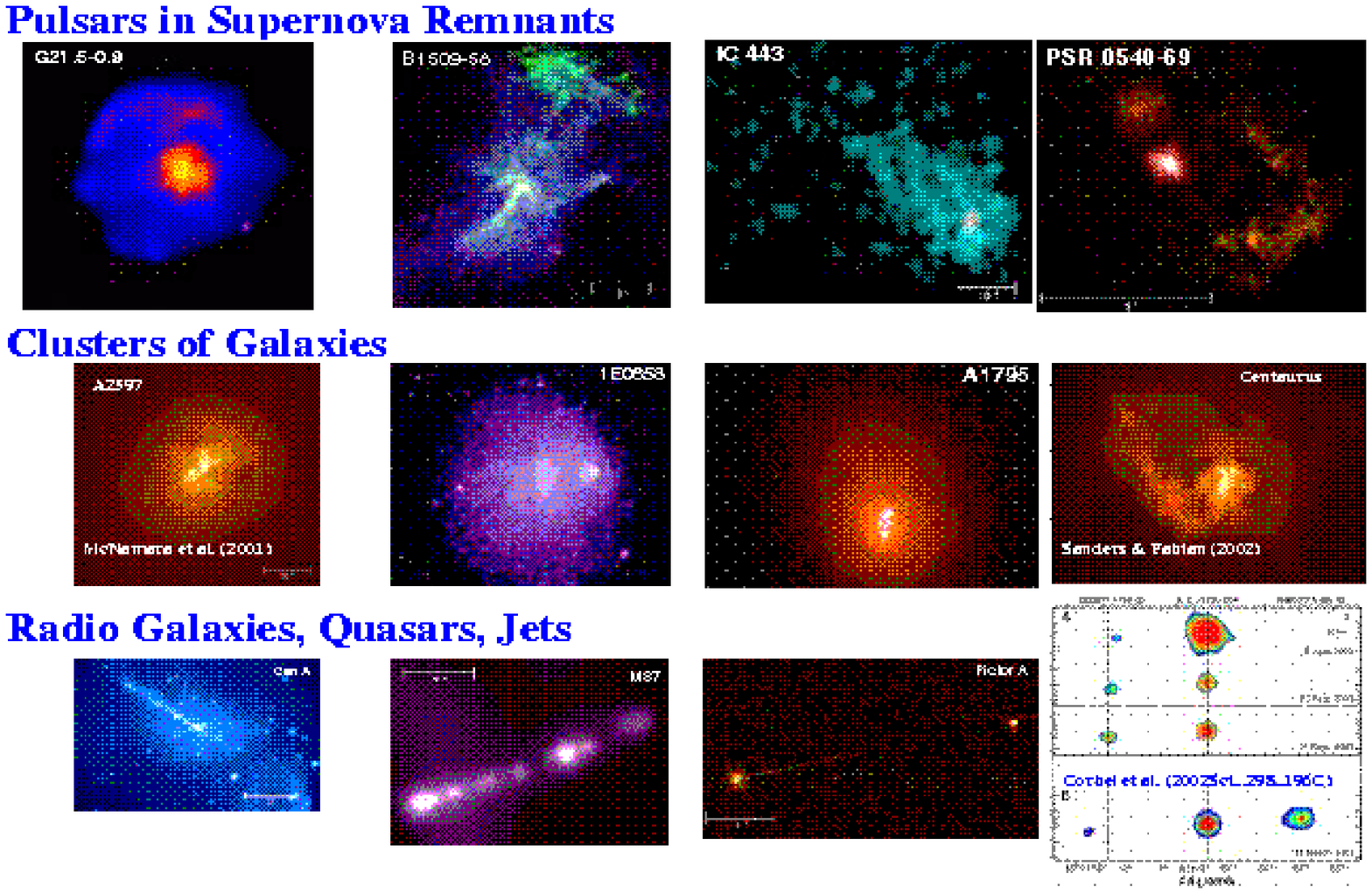}
\caption{\emph{Chandra} images of examples of the systems discussed in
this review. Credits: NASA/MSFC/SAO/CXC (top row, middle row, left 3
on bottom row);  \cite{Corbel02} (bottom row right).
\label{introImages}} 
\end{figure}

Figure \ref{introImages} provides examples of systems representing
the three topics I will discuss: pulsars and supernova remnants; cooling 
flows in clusters of galaxies; and jets in quasars and radio galaxies.

In clusters of galaxies, the X-ray emission from the hot gas filling
the volume between the galaxies was shown to have temperatures
consistent with the gravitational potential of the cluster. In a
substantial fraction of clusters, the gas was observed to be
sufficiently dense that it would cool in much less than a Hubble time,
and it was interpreted  that massive cooling flows involving hundreds of
solar mass per year were condensing onto the cluster centers
(\cite{Arnaud88}, \cite{Fabian94}). This 
created a great puzzle, as neither the destination of the cooling gas,
nor great quantities of gas at temperatures less than 1 to 2 keV have
been found.  It now appears that powerful radio sources in cD galaxies
in the cluster cores (\cite{Burns90}) provide the energy to offset the
cooling.  

In contrast to accretion, a major theme of radio astronomy has been
the origin of cosmic rays, the acceleration of particles to ultra high
energies, and the transport of energy in jets to distances of pc to Mpc
away from the nuclei of active galaxies. Radio astronomy has been primarily
an imaging rather than spectroscopic science (with some important
exceptions which we have heard at this symposium). In this article I
therefore emphasize X-ray imaging. For decades radio observations have
studied detailed structure in supernova remnants, emission from
cluster of galaxies and sources in clusters, and jets in active
galaxies. With the half-arcsecond X-ray imaging now available from the
\emph{Chandra} X-ray Observatory, we can finally compare X-ray with
GHz radio observations on the same angular scales.

 We adopt a flat,
accelerating cosmology with Hubble constant H$_{0}$=70
km s$^{-1}\, \rm{Mpc,}^{-1} \Omega_{m}$=0.3, and
$\Omega_{\Lambda}$=0.7.

\section{Pulsars and Supernova Remnants \label{pulsars}}

\begin{figure}[b]
\includegraphics[viewport=146 4  335 508,clip,angle=-90.,width=5.in]{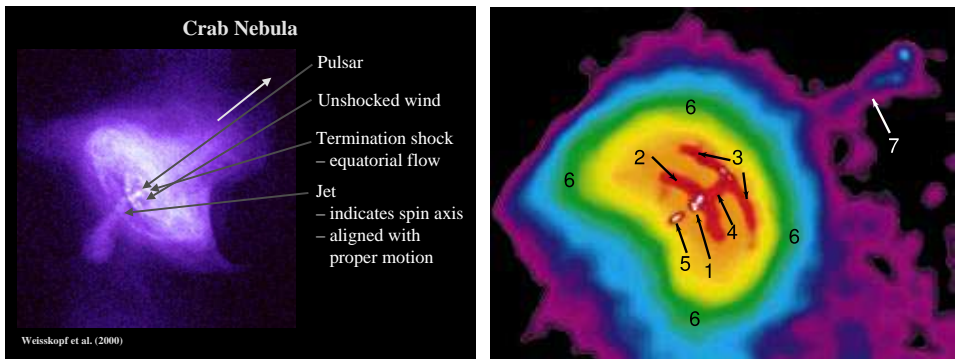}
\caption{\emph{Chandra} images of the Crab Nebula (left, NASA/CXC, from
\cite{Weisskopf00}) and the Vela Pulsar Nebula (right, from
\cite{Pavlov03}). The large arrow on the left shows the direction of
the spin axis and the proper motion, for \emph{both} pulsars.
Numbered regions in Vela indicate 1. pulsar; 2. termination shock;
4. and 5. the inner jet and counterjet, respectively; 7. the outer
jet.  \label{crabvela}}
\end{figure}

From the earliest radio observations of pulsar periods, and of the rates 
at which the periods were increasing, it was found that from 10$^{35}$ to
10$^{39}$ erg s$^{-1}$ of rotational energy was being lost.  The radio 
luminosities were typically orders of magnitude smaller,  and even in those
cases with X-ray emission, such as the Crab Nebula and Vela Pulsar (Fig.
\ref{crabvela}),  the total radiated luminosity was a very small fraction
of the energy loss. It was presumed that the balance went into
relativistic particles (\cite{Michel69a}) and driving a wind (\cite{Michel69b}).  In some cases, a plerion, or pulsar wind nebula,
was present, and interpreted as the manifestation of the outflow. The 
picture was that a cold, relativistic gas was driven away from the speed of
light cylinder.  This gas would not be directly detectable, until it 
formed a shock where it encountered the nebular gas. \cite{Gaensler03}
gives an excellent review of all these issues.

We now see this 
termination shock in the Crab and Vela as indicated in Figure \ref{crabvela}.
The figure shows rings indicating equatorial flow, and jets. The jets must 
be along the rotational axis, and not the magnetic pole, since the latter 
must sweep a wide cone on the sky, giving pulses when it intersects our
line of sight. These pulsars give us a clear case of a jet which is 
associated with rotation.

\begin{figure}[h]
\includegraphics[
width=.96\textwidth]{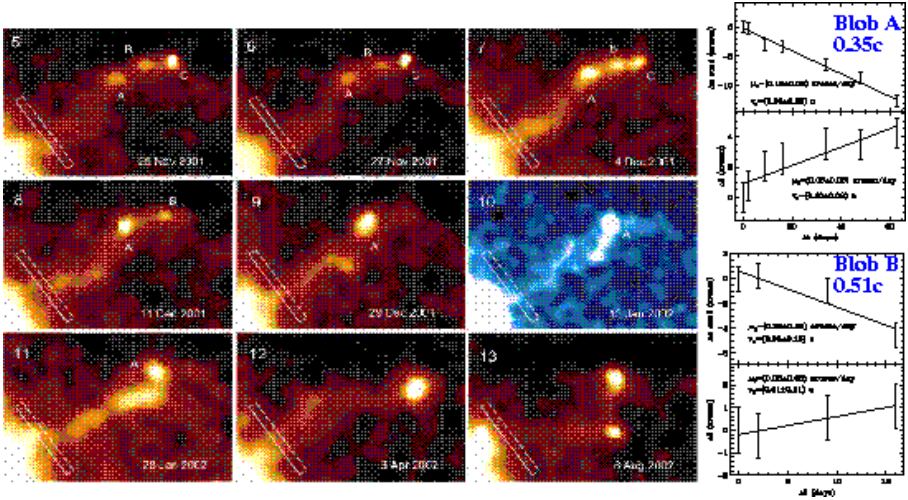}
\caption{Time sequence of the outer Vela jet (\cite{Pavlov03}),
showing remarkable changes in morphology, and apparent relativistic
motion of blobs. Image 10 is with the HRC, the others are with ACIS
S3. \label{velajet}}
\end{figure}

Figure \ref{velajet} (from \cite{Pavlov03}), shows a sequence of X-ray images
of the outer jet in the Vela Pulsar Nebula, over an 8 month time interval.
There are distinct morphological and brightness changes, e.g., in as few as
7 days from 4 to 11 December, 2001.  These changes are attributed to 
magnetic instabilities in the jet, and/or its interaction with the 
interstellar medium.   

Figure \ref{velajet} (right column) plots the position vs. time for
two of the knots, ``A'' and ``B'' relative to the narrow rectangle
which is superposed in fixed celestial coordinates in the images. At
the distance of the Vela pulsar, 300 pc, these motions respectively
correspond to 0.35 and 0.51 times the velocity of light. The changing
image may be due to  moving  material or to moving patterns in  the jet.

\begin{figure}[h]
\includegraphics[
width=\textwidth]{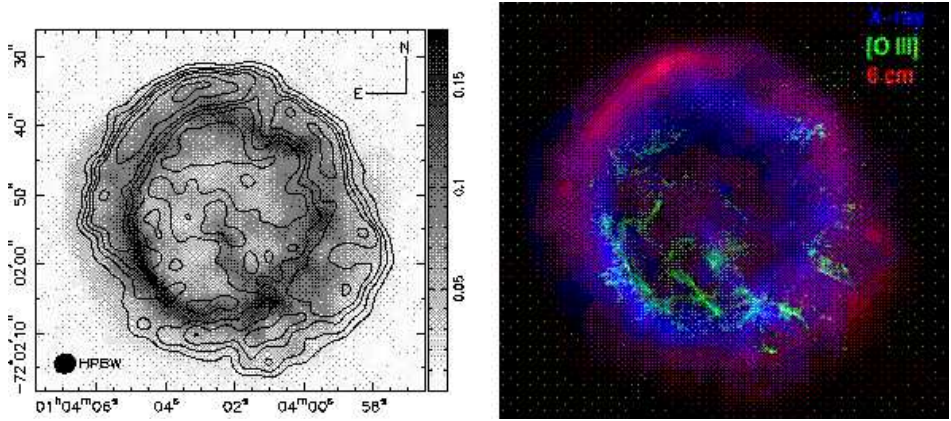}
\caption{Left Panel (Figure 5 from \cite{Gaetz00}): X-ray (gray scale) and radio (contours) images of
  the supernova remnant E0102-72 in the SMC. Right Panel (Figure 6 from \cite{Gaetz00}): X-ray
  (blue), 6 cm (red) and optical [O III] emission (green). \label{E0102}}
\end{figure}

Figure \ref{E0102} (from \cite{Gaetz00}, figs 5. and 6.) shows the
supernova remnant 1E 0102.2-7219 in the Small Magellanic Cloud (SMC).
It is a young, oxygen rich remnant, which has been observed repeatedly
since orbital activation as a calibration target. \cite{Flanagan00},
and \cite{Fredericks01} discuss HETG observations which show details
of the X-ray line ionization structure and velocities.  \cite{Gaetz00}
discuss how the X-ray picture shows for the first time the sharp
interaction of the the outgoing shock with the interstellar medium,
delineated by the outer edge of the X-ray gray-scale image to the
left. The most intense X-ray ring is the reverse shock, moving back
into the expanding medium. The radio emission is brightest in the
region between the two shocks, with the interpretation that
the shocks are intensifying the magnetic fields, and accelerating
the particles.

\begin{figure}[h]
\includegraphics[
width=.96\textwidth]{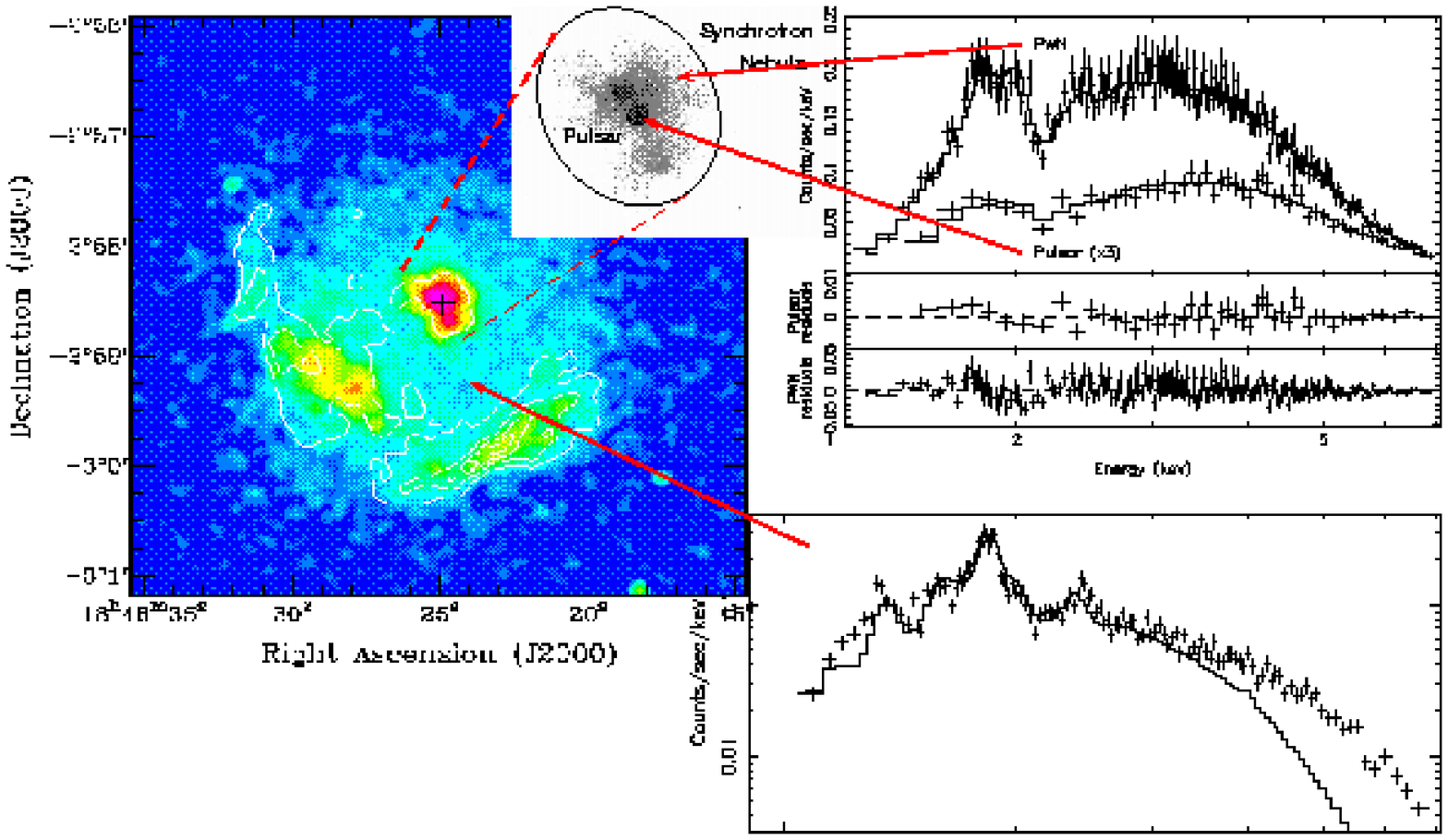}
\caption{The composite supernova remnant Kestevan 75 (adapted from
  Figures 1, 2, 3, and 7 of \cite{Helfand03}). Left panel:
  False color X-ray image with 20 cm
  radio contours. Top center: X-ray image of the
  pulsar and synchrotron nebula. Right panels:  X-ray
  spectra of the pulsar, synchrotron nebula, and a region of the shell
  inside the shock. \label{KES 75}}
\end{figure}

The composite supernova remnant Kes75 is
discussed by \cite{Helfand03} (figure
\ref{KES 75}). The false-color X-ray image in the left panel shows the
shell structure of the outer shock, the peak X-ray intensity
consistent with a reverse shock, and the 20 cm VLA radio emission from
the outer, shocked X-ray emitting regions, all as in E0102-72.  There
is also a radio and X-ray bright central nebula, for which their
expanded image (top center) shows a synchrotron nebula, a pulsar, and
hot spots to the NNE and SSW which might be from a
jet. \cite{Helfand03} analyze the spectrum of the pulsar, and the
nebula, separately, and show they are continuous, non-thermal
emission, which they attribute to synchrotron radiation (figure
\ref{KES 75}, top right panel). Perhaps their most interesting
discovery is in the X-ray spectrum of a broad region of the shell
remnant,  shown in the lower right.  While lines of Mg, Si, and S are
clearly present, and require a hot gas, no purely thermal emission model
(indicated by the histogram) can provide both the line emission and
all the continuum emission up to the 7 keV plotted in the figure. They
model a combination of a thermal and a synchrotron spectrum, where the
latter is due to a dust halo scattering the X-ray emission from the
pulsar and its synchrotron nebula.

\section{Cooling Flows in Clusters of Galaxies \label{cool}}

\begin{figure}[h]
\includegraphics[
width=.94\textwidth]{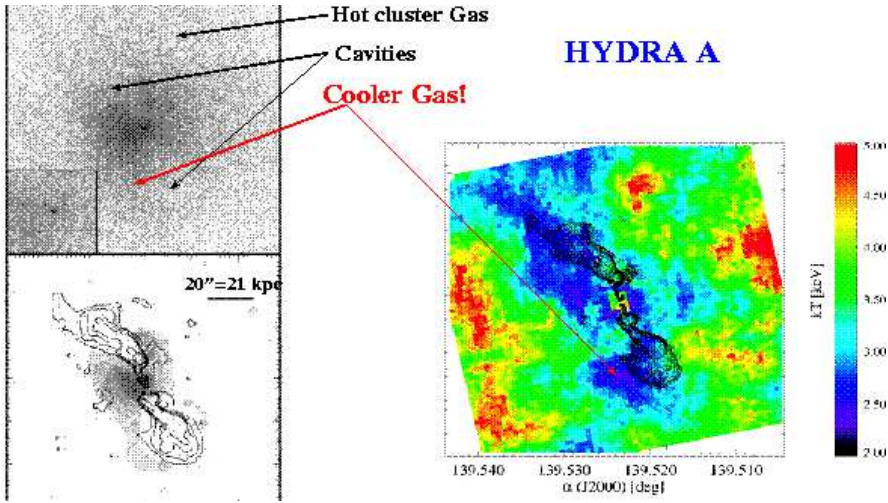}
\caption{X-ray (gray scale and false color) image and 6 cm contours of
  the central region of the Hydra A cluster. (Left, from
  \cite{Mcnamara00}; right from \cite{Nulsen02}.) \label{hydraA}}
\end{figure}

The \emph{Chandra} calibration observation of the Hydra A cluster of
galaxies is shown in figure \ref{hydraA} (from \cite{Mcnamara00} and
\cite{Nulsen02}).  Hydra A had been inferred to have a large cooling
flow, of order 250 solar masses per year (\cite{White97}). However,
only of order 10 solar mass per year could be accounted for by
star formation rates (\cite{Mcnamara95}, \cite{Hansen95}). A general
suggestion to prevent cooling flows condensing in cluster centers  was
that radio sources in the central cD galaxy provided the energy to
balance the radiation loss (cf. \cite{Tucker83}). A convential picture was that the radio
plasma was emitted in a jet, which generated a shock upon colliding
with the cluster gas (\cite{Clarke97}, \cite{Heinz98}). 

The left panel of Figure \ref{hydraA} was discussed by
\cite{Mcnamara00}, who emphasized the coincidence of the apparent
``cavity'' in the X-ray emission (gray scales in top and bottom
panels) with the 6 cm VLA emission (contours in bottom panel, from
\cite{Taylor90}). However, as shown in the temperature map of the
cluster by \cite{Nulsen02}, (right panel), the higher density gas
surrounding the cavities is at a \emph{lower} temperature than the
more distant cluster gas, and thus is not the result of a strong
shock.  This leads to the interpretation that the ``cavity'' is
actually a buoyantly rising ``bubble.''

We can use the X-ray observations to learn much about the conditions
in cluster cores in general.  We can measure, as a function of
position, the electron temperature, T, from the shape of the X-ray
spectrum, and the density n$_{\rm {e}}$ from the measured intensity,
which is proportional to n$_{\rm {e}}^2 \sqrt{\rm {T}}$. The
coincidence of the X-ray cavities with the radio emission gives us
three more relations: The work, p$\Delta$V, done to create the cavity
should equal the energy available from the radio jet; the pressure of
the radio plasma should approximately balance the X-ray gas pressure
at the cavity boundaries; and the time scale for the bubble to rise
and expand, gives lifetimes. In particular, both the observation that
the bubbles do not expand faster than the sound speed (due to lack of
strong shocks), and the estimate that they rise with a velocity of
order the Keplarian velocity at their position (\cite{Churazov00}),
leads to velocities of order hundreds of km/s, and ages of a few
10$^7$ years.

In the case of Hydra A, \cite{Nulsen02} estimate p$\Delta$V $\approx
1.2 \times 10^{59}$ ergs, while if one assumes the magnetic field and
particles in the radio plasma have a minimum energy density (which is
very nearly the equipartition value) this density times the
volume is only u$_{me}\,{\rm {V}}\,\approx 1.4 \times 10^{58}$ ergs. On
the other hand, based on the rotation measure, \cite{Taylor90} argue
that the magnetic field is several times less than the equipartition
value, and if so the energy in the radio plasma could be comparable to
the work done creating the cavities.

\begin{figure}[ht]
\includegraphics[
width=.96\textwidth]
{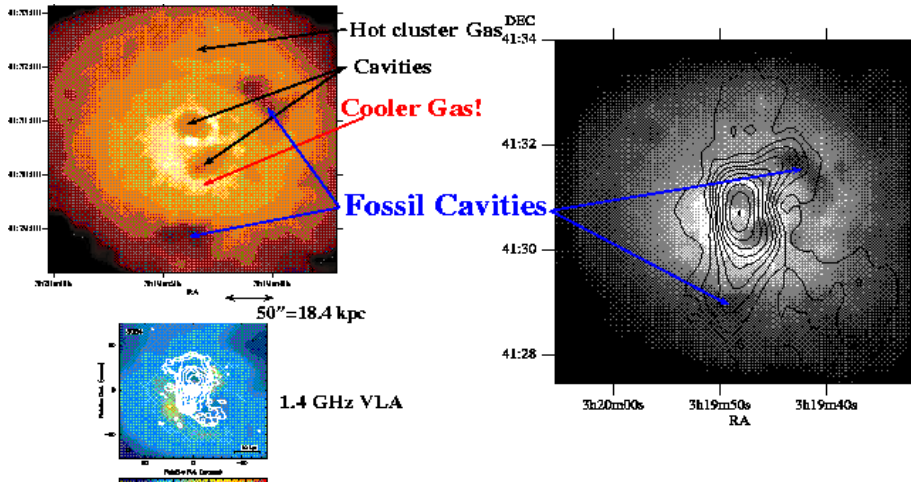}
\caption{Left panels (from \cite{Fabian00}, figs. 5 and 7): False
  color X-ray image of the center of the Perseus cluster, showing
  the inner X-ray cavities filled with 1.4 GHz emission. Right panel
  (from \cite{Fabian02}, Fig. 3): gray scale X-ray image, showing the
  74 MHz radio contours extending to the outer ``fossil'' cavities. \label{perseusA}}
\end{figure}

The remarkable original discovery of cluster gas cavities coincident
with radio emission was made by \cite{Bohringer93} using \emph{ROSAT}
high resolution imager observations of the Perseus Cluster. This
cluster contains the powerful radio source 3C 84.  The original
\emph{Chandra} observations are shown in Figure \ref{perseusA}, from
\cite{Fabian00} (left panels) and \cite{Fabian02} (right panels).  We
see the original cavities, (also called ``holes''), within about 10
kpc nearly  N and S of the cluster center.  \emph{Chandra} data
shows that the X-ray bright rims of the holes are \emph{cooler}, about
2.7 keV, than the extended cluster emission which is at about 6.5 keV
(\cite{Fabian00}). As in the case of Hydra A, the cavities therefore could not
be interpreted as shock heating by the radio plasma as predicted, e.g.
by \cite{Heinz98}.

 The work to evacuate the North cavity is about p $\Delta$V
$\approx 2 \times 10^{58}$ ergs. A standard minimum energy
calculation, which takes a filling factor $\phi$=1 and a ratio of
proton to electron energy density k=0, gives u$_{me} \,\Delta V
\approx 4 \times 10^{56}$ ergs, almost a factor of 100
smaller. \cite{Fabian02} use this to infer that the ratio $\frac{\rm
{k}+1}{\phi}$ is of order 600, and at least 180 to establish pressure
equality, all under conditions of minimum energy.  On the contrary, in
order that the synchrotron lifetime of the electrons emitting at 1.4
GHz be longer than the age of the N cavity, \cite{Fabian02} find
that the magnetic field is 4 to 10 times less than equipartition
value, (which also requires $\frac{\rm {k}+1}{\phi} < 500$).

The regions of lower surface brightness 1\farcm5 to the NW and 
2\arcmin\ to the S, are interpreted by \cite{Fabian02} as due to
bubbles for which the high frequency emission is absent due to
ageing of the radio electrons. They argue that the 74 MHz radio images
(\cite{Blundell02}) which point to these ``fossil'' or ``relic''
cavities, trace earlier radio jets which evacuated and energized these
regions. 

The mechanism by which the central, active radio source provides the
energy to balance the cooling flow may have been seen directly in a
200 ks observation of Perseus A (\cite{Fabian03}). They explore the
possibility that faint X-ray intensity ripples (seen in their Figure
3.) are sound waves moving out from the interaction of the jets
producing the bubbles. Since they do not detect any temperature
variations in these regions, density oscillations most economically
explains the intensity variations.  They invoke viscosity to dissipate
the sound energy into the required heating of the gas.

The lesson of all the bubbles in X-ray clusters (e.g.,
\cite{Blanton01}, \cite{Mcnamara01}, \cite{Sanders02}, \cite{Sun03})
is that the central black hole must be producing much more energy than
we see from the direct radio emission or from the minimum energy
arguments.  It is therefore likely that a central active galaxy
produces the energy output needed to counteract the cooling flows.
\cite{Nulsen03} argues that this proceeds via a feedback process,
whereby the central AGN becomes more active whenever the cooling gas
can effectively accrete to the central black hole, and the increased
activity then retards or stops the cooling flow.

\section{Jets in Radio Galaxies and Quasars \label{jetsection}}

X-ray jets in a diverse collection of objects has become possible as a
distinct field of research only with the 100-fold improvement in
2-dimensional imaging afforded by the 0\farcs5 resolution of
\emph{Chandra}, compared to the previous 5\arcsec\ imaging of the
\emph{Einstein} and \emph{ROSAT} observatories. Jets are studied in
contexts as varied as the symbiotic binary R Aqr (\cite{Kellogg01}),
the galactic black hole candidate XTE~J1550-564 (\cite{Corbel02}),
plerions (discussed in Section \ref{pulsars}), low power FR I sources
where the X-ray emission is interpreted as an extension of the radio
synchrotron emission (e.g., \cite{Worrall01}), and in powerful FR II
galaxies and quasars.  This section will consider only the last topic.

\begin{figure}[h]
\includegraphics[
width=.96\textwidth]{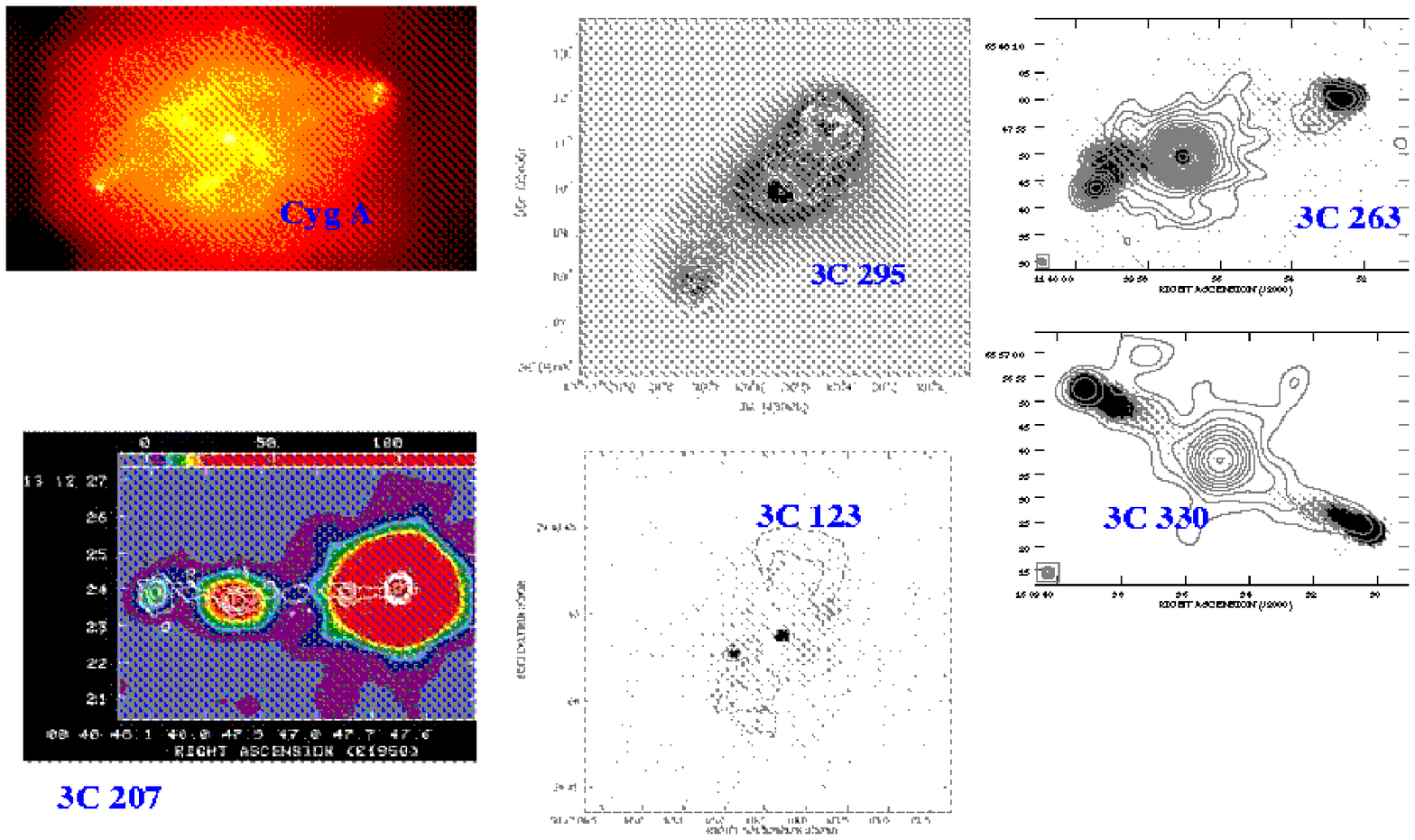}
\caption{X-ray hotspots (false color and gray scale) at the end of
  radio jets (contours). Clockwise from upper left: Cygnus A (from
  \cite{Wilson00}), 3C 295 (from \cite{Harris00}), 3C 263 and 3C 330
  from \cite{Hardcastle02}), 3C 123 (from \cite{Hardcastle01}), and 3C 207
  (from \cite{Brunetti02}).
\label{hotspots}}
\end{figure}

In Figure \ref{hotspots} I select some examples which illustrate the
X-ray emission at the hotspots at the end of jets. In all these cases,
from the radio observation of flux and spatial extent, plus the
assumption of equipartition, one can conclude that the synchrotron
photon density is the largest energy density in the X-ray emitting
region. Thus the natural X-ray mechanism is synchrotron self-Compton
(SSC) emission. The X-ray fluxes are generally consistent with
magnetic fields in the range 70 to 320 $\mu$Gauss, which are just a little
below the equipartition field values (see refs. given in figure
caption). This is a body of evidence that conditions near
equipartition might be a reasonable assumption.

\begin{figure}[ht]
\includegraphics[
width=.96\textwidth]{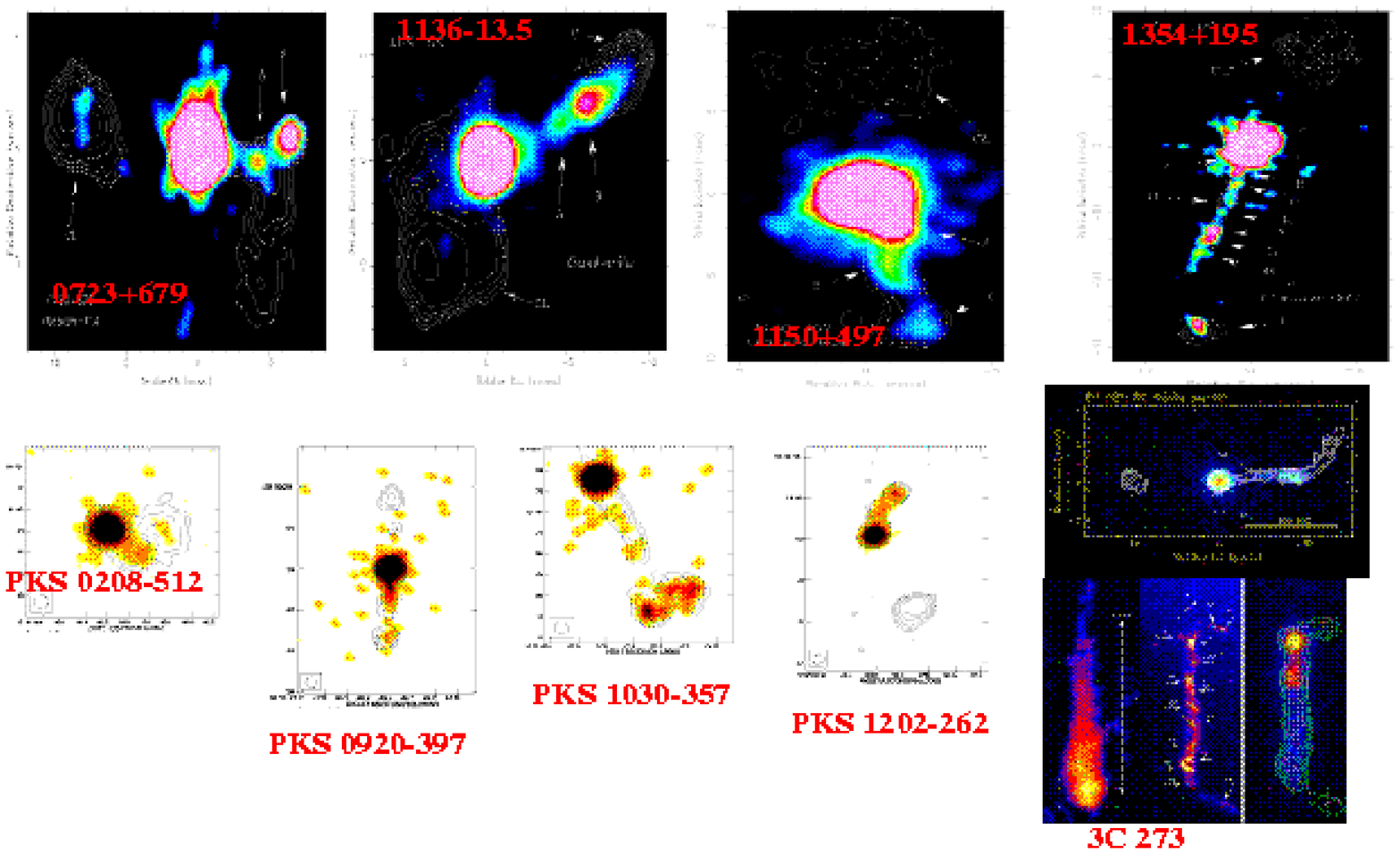}
\caption{Chandra images of X-ray jets. Top panel from
  \cite{Sambruna02}. Bottom: PKS objects from \cite{Schwartz03b};
  PKS~0637-752 from \cite{Schwartz00}; 3C 273 from \cite{Marshall01} \label{jets}}
\end{figure}

Figure \ref{jets} shows some of the X-ray images of jets (in false
color), overlaid with radio contours. These are samples from two
surveys, one by \cite{Sambruna02}, the other by
\cite{Marshall02}, (also reported in \cite{Schwartz03}). We also see the
discovery image of PKS~0637-752 (\cite{Schwartz00}, and a
radio-optical-X-ray view of the jet in 3C 273 (\cite{Marshall01}).
The X-ray emitting regions closely follow the radio in general, but
the intensities sometimes correlate closely, as in the straight western jet
of PKS~0637-752, and sometimes anticorrelate as in 3C~273. We will use the
four PKS objects from our survey (bottom of Figure \ref{jets}) to
discuss the physical conditions in jets (\cite{Schwartz03b}).

\begin{figure}[h]
\includegraphics[
width=5.in]
{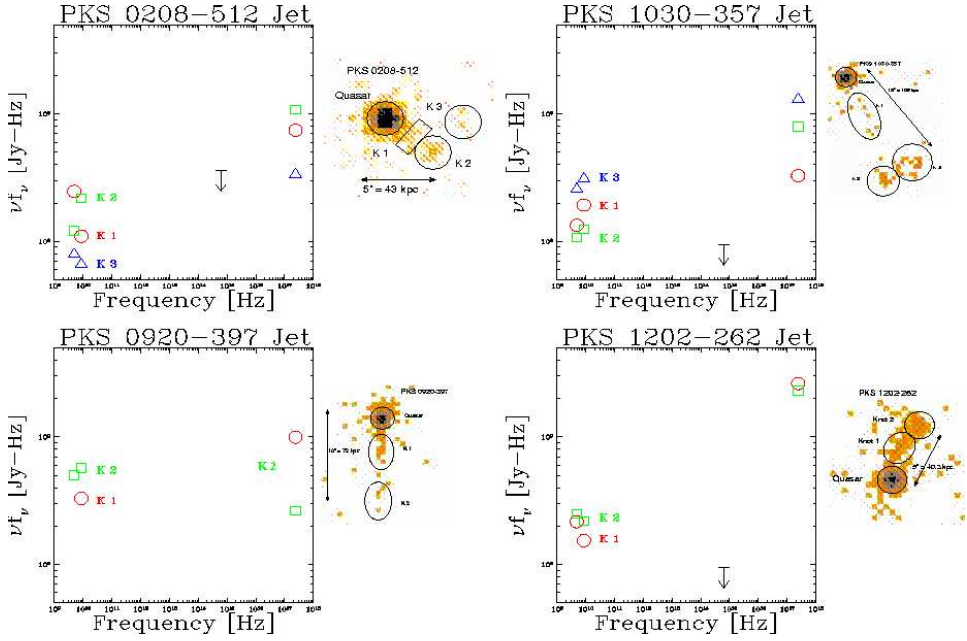}
\caption{Spectral energy distributions of regions within the X-ray jets. \label{jetSED}}
\end{figure}

We have 8.64 GHz ATCA images of all our sources, and in some cases 4.8 GHz
data. We smooth both the radio and X-ray to a 1\farcs2 resolution, and
superpose them by forcing coincidence of the quasar cores. We then
divide the jets into distinct regions.  This is somewhat subjective,
guided by the features in the radio and X-ray emission. To label
the different regions we use the term ``knots'' (K), with numbers
increasing away from the quasar, but we don't intend this to prejudice
the nature of the actual structure. 

Figure \ref{jetSED} shows the spectral energy distributions we
construct for each region. Optical upper limits from Magellan
observations (J. Gelbord, private communication and in preparation
2004) directly show that for most cases the X-ray emission cannot be a
simple extension of the radio synchrotron spectrum. For other regions,
e.g., K3 of PKS~0208-512, the radio spectral shape would not connect
to the X-ray region. For some other regions, e.g., K2 of PKS~0920-397
and K2 of PKS~0208-512 the X-ray emission could well result from a
continuation of the relativistic electrons to high enough energy to
emit X-ray synchrotron emission, as inferred for the first knot, A1,
in 3C 273 (\cite{Marshall01}).  The simplest X-ray emission mechanism,
given the strong correlation with the radio, should invoke radiation
from the same spectrum of relativistic electrons. This indicates some
form of inverse Compton (IC) emission.  From the size and radio emission,
we know that SSC will not be important, and at 10's to 100's of kpc
from the quasar the energy density of photons from the central black
hole will not give significant radiation.  

The most likely target photons for IC emission are the cosmic
microwave background (CMB). This was originally discussed by
\cite{Felten66} in the context of explaining the cosmic X-ray
background. However, in the original case of PKS~0637-752, calculating
the equipartition magnetic field gave a result about 100 times larger
than the maximum magnetic field which would allow the X-rays to be
produced by IC/CMB radiation.  The problem of requiring total energies
more than 10$^3$ times larger was resolved by \cite{Tavecchio00} and
\cite{Celotti01}, who considered the enhancement of the apparent CMB
density by the factor $\Gamma^2$, (\cite{Dermer94}), in a frame moving
with bulk relativistic velocity $\beta = \sqrt(1-1/\Gamma^2)$ with
respect to the CMB frame. If one plots the required relativistic
beaming factor $\delta=(\Gamma(1-\beta \cos\theta))^{-1}$ against the
required rest frame magnetic field, then since $\delta \propto \rm
{B_{IC}}$  and $\delta \propto 1/\rm {B_{eq}}$, one can always find a
solution for $\delta$ and B for which the source is near equipartition
in its rest frame, and the same population of electrons produces radio
synchrotron radiation in the B field, and X-ray inverse Compton
radiation off the cosmic microwave background.

\begin{figure}[h]
\includegraphics[
width=.96\textwidth]
{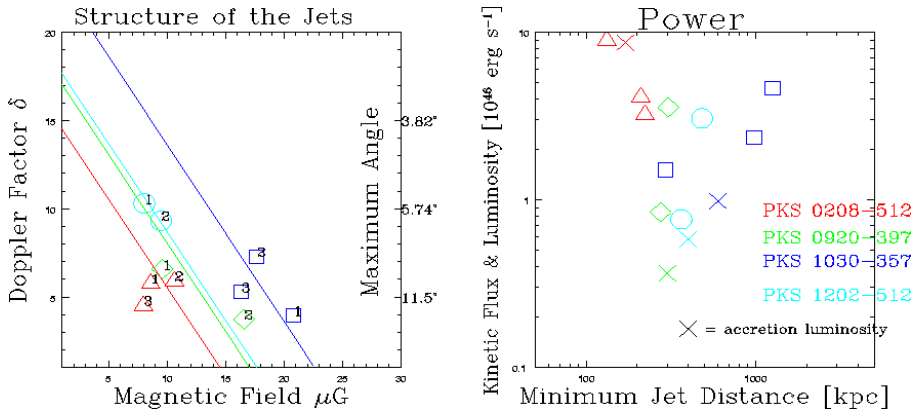}
\caption{Structure of the X-ray jets. Left panel shows the Doppler
  factors and rest frame magnetic fields inferred for each
  region. Uncertainties  are systematics dominated and about a factor
  of 2. Solid lines show the loci of constant kinetic flux. Right
  panel plots the kinetic flux through each region, vs. the minimum
  space distance of each component. Colored triangles, diamonds,
  circles and squares indicate the same source in each frame. Crosses
  plot the bolometric radiative luminosity of the quasar.  \label{jetstruct}}
\end{figure}

We will continue by assuming all the jet emission is due to
IC/CMB. Figure \ref{jetstruct} (left panel) shows the range of Doppler
factors, $\delta \approx$ 2 to 10, and intrinsic magnetic fields, B
$\approx$ 5 to 25 $\mu$G, for these objects
(\cite{Schwartz03b}). In this derivation, we have had to assume that
$\Gamma = \delta$, since we do not have independent information on the
orientation of the jet.  From $\delta$, we can infer a
maximum angle of the jet from our line of sight, $\theta_{\rm {max}} =
\cos^{-1}[(\delta -1/\delta)/\sqrt(\delta^2-1)]$, which we use to
compute the minimum space distance (3 dimensions) of each region from
the central quasar. We estimate (right panel of Figure
\ref{jetstruct}) the kinetic flux of each element as A$\Gamma^2$cU
(e.g., \cite{Ghisellini01}), where A is the cross sectional area (and
we assume cylindrical symmetry), and U the total energy density in the
rest frame of the jet.  We assume equipartition, with an equal energy
in protons and electrons, so that U=3 B$^2$/(8 $\pi$). Under all these
assumptions, the lines $\delta \propto $ 1/B are the lines of constant
kinetic flux, as shown in the left panel.

In the right panel, the crosses plot the bolometric radiative
luminosity of the quasar cores. We see that the kinetic flux in the
jet is comparable to or greater than the accretion flux. This is
consistent with the conclusion of \cite{Meier03} that accretion flow
models must also consider jet production.

One of the most dramatic implications of the inference that we see
IC/CMB X-radiation from radio jets  is that any given object would
appear to have a constant X-ray surface brightness even as it were
displaced to an arbitrarily large redshift (\cite{Schwartz02}). This
is because the energy density of the CMB increases as (1+z)$^4$,
exactly offsetting the (1+z)$^{-4}$ cosmological diminution of surface
brightness. Since the observed X-ray jet structures have 
length scales of 10's of kpc projected on the sky, they will be
at least several arcsec long at redshifts greater than 2, and would
easily be resolved by Chandra.  In fact, all objects intrinsically
similar to PKS~0637-752, or the outer knots of 3C~273, would be bright
enough to already be detected by \emph{ROSAT}, but would appear as
point sources to the resolution of the PSPC all-sky survey.

Where are these bright X-ray jets at high redshifts? They could not be
recognized as extended in the \emph{ROSAT} all sky survey, so they
would most likely be identified simply as part of the quasar core
emission.  If the quasar itself were not recognized, e.g., because it
was too faint, the jet could be among the miscellaneous unidentified
sources.  Alternately, the jet could outshine the quasar in X-rays,
and be cataloged at a position some distance away from the quasar, and
again be a  miscellaneous unidentified source.

\begin{figure}[h]
\includegraphics[
width=.96\textwidth]
{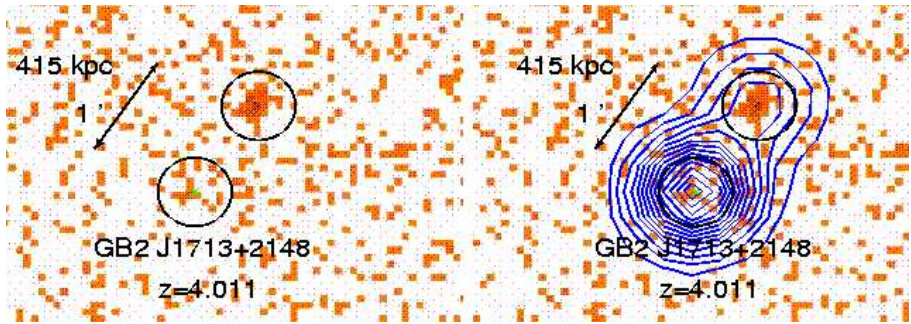}
\caption{Left panel shows a \emph{ROSAT} HRI observation of the quasar
  GB 1713+2148 (green cross). The quasar is at most a  3$\sigma$
  detection in the 20\arcsec\ extraction circle. A stronger,
  unidentified source lies $\approx$ 1\arcmin\ to the NW. The right
  panel superposed the NVSS 1.4 GHz radio contours, clearly showing
  the connection of the stronger source to the quasar.  \label{jet1713}}
\end{figure}

Figure \ref{jet1713} shows a possible example of the last case. In
this \emph{ROSAT} HRI pointed observation, there was only a possible
3$\sigma$ detection of the quasar GB 1713+2148 (\cite{Vignali03}).
There is an obvious stronger, point source about 1 arcmin to the NW
(left panel of Figure \ref{jet1713}).  When the NVSS 1.4 GHz image
(\cite{Condon98}) is superposed, we see the contours of the radio loud
quasar extending around this source (\cite{Gurvits03}), so that it is
clearly associated with the quasar, at a redshift z=4.011
(\cite{Hook98}). We still need a high resolution X-ray image to
ascertain if this is really an X-ray jet, or perhaps just a hotspot or
lobe.

\begin{figure}[h]
\includegraphics[viewport=190 120  420 747,clip,
angle=-90,width=.96\textwidth]
{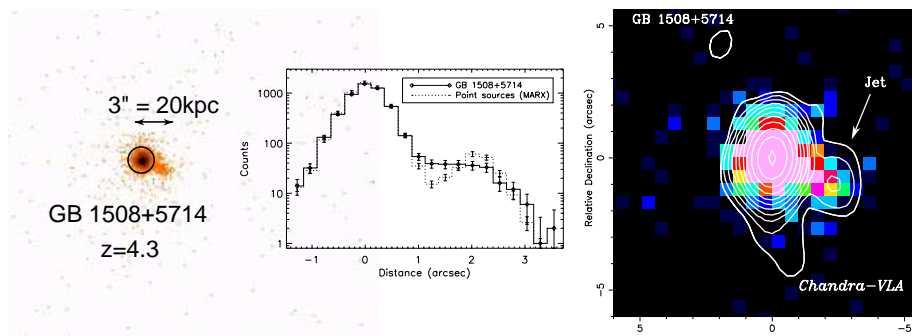}
\caption{Left and center from \cite{Siemiginowska03}: \emph{Chandra} image binned in 0\farcs15 pixels, clearly
  showing extent to the WSW of the quasar. The circle is 2\farcs5
  diameter, and contains 95\% encircled power. The solid histogram
  is the intensity profile along the jet, which clearly cannot be
  simulated by two point sources (dotted histogram). Right:
  \cite{Cheung04} subsequently discovered a radio jet in archival 1.4
  GHz VLA data. \label{jet1508}}
\end{figure}

Another case is due to \cite{Siemiginowska03} (Figure
\ref{jet1508}). Here the X-ray image (left panel) is clearly extended,
and the analysis in the central panel shows that the data (solid line)
cannot be simulated simply by two point sources (dotted
line). \cite{Cheung04} has analyzed archival VLA data, and found a
coincident jet at 1.4 GHz (right panel).  An IC/CMB analysis shows
that this jet must be in relativistic motion, with a Doppler factor
$\delta \ge$2.6 and a magnetic field B$\le$161$\mu$G, where the
uncertainty is largely due to the uncertain slope of the radio and
X-ray emission (\cite{Siemiginowska03}).

\section{Summary}

We have reviewed some examples where X-ray and radio astronomy act in
conjunction to reveal the workings of energy and matter outflow. Radio
observations show the presence of magnetic fields, and allow
calculations of \emph{minimum} energy densities and total
energy. X-ray observations can sometimes break the degeneracy between
the magnetic field and particle densities and thus test the
equipartition assumptions, although notably in the case of quasar jets
the X-ray observations add a new parameter by revealing bulk
relativistic motions.  In the case of clusters of galaxies, the radio
observations apparently have solved the problem of retarding the
cooling flows in clusters of galaxies, while the energetics required
by the X-ray data lead to an inference of much larger total energies
being supplied by the central black hole. 

\vspace{0.1in}

\textbf{Acknowledgements} 

This work was supported in part by NASA contract NAS8-39073 to the
Chandra X-ray Center, and NASA grant GO2-3151C to SAO. This research
used the NASA Astrophysics Data System Bibliographic Services, and the
NASA/IPAC Extragalactic Database (NED) which is operated by the Jet
Propulsion Laboratory, California Institute of Technology, under
contract with the National Aeronautics and Space Administration.



\begin{thebibliography}{}

\bibitem[Arnaud (1988)]{Arnaud88} Arnaud, K. A. 1988, in ``Cooling
  flows in clusters and galaxies'', Proceedings of the NATO Advanced
  Research Workshop, (Dordrecht:  Kluwer Academic Publishers), 31

\bibitem[Atoyan \& Dermer (2001)]{Atoyan01} Atoyan, A., \& Dermer,
C. D. 2001, PRL, 22, 1102

\bibitem[Blanton \etal\ (2001)]{Blanton01} Blanton, E. L., Sarazin, C. L.,
  McNamara, B. R., \& Wise, M. W. 2001, ApJ, 558, L15

\bibitem[Blundell \etal\ (2002)]{Blundell02} Blundell, K. M., Kassim,
  N. E., \& Perley, R. A. 2002, in Rao, A. P., Swarup, G., \&
  Gopal-Krishna, eds., Proc IAU Coll. 199, ``The Universe at Low Radio
  Frequencies,'' 189, (astro-ph/0004005)

\bibitem[B\"{o}hringer \etal\ (1993)]{Bohringer93}B\"{o}hringer, H.,
 Voges, W.,  Fabian, A. C., Edge, A. C., \&  Neumann, D. M. 1993,
 MNRAS, 264, L25

\bibitem[Bowyer \etal\ 1970]{Bowyer70} Bowyer, C. S., Lampton, M.,
 Mack, J., \&  de Mendonca, F. S.   1971, ApJ, 161, L1

\bibitem[Bradt \etal\ 1967]{Bradt67} Bradt, H., Mayer, W., Naranan, S.,
 Rappaport, S., \& Spada, G. 1967, ApJ, 150, L199

\bibitem[Brunetti \etal\ (2002)]{Brunetti02} Brunetti, G., Bondi, M.,
  Comastri, A., \& Setti, G. 2002, A\& A, 381, 795

\bibitem[Burns (1990)]{Burns90} Burns, J. O. 1990, AJ, 99, 14

\bibitem[Celotti \etal\ (2001)]{Celotti01} Celotti, A., Ghisellini,
  G., \& Chiaberge, M. 2001, MNRAS, 321, L1.

\bibitem[Cheung (2003)]{Cheung04} Cheung, C. C. 2004,   ApJ, 600, L23

\bibitem[Churazov \etal\ (2000)]{Churazov00} Churazov, E., Forman, W.,
 Jones, C., \& Böhringer, H. 2000, A\&A, 356, 788

\bibitem[Clarke \etal\ (1997)]{Clarke97} Clarke, D. A., Harris, D. E.,
  \& Carilli, C. L. 1997, MNRAS, 284, 981

\bibitem[Condon \etal\ (1998)]{Condon98} Condon, J. J., Cotton, W. D.,
 Greisen, E. W., Yin, Q. F., Perley, R. A., Taylor, G. B., \&
 Broderick, J. J. 1998, AJ, 115, 1693 

\bibitem[Corbel \etal\ 2002]{Corbel02} Corbel, S. \etal\ 2002, Science, 298, 
196

\bibitem[Dermer \& Schlickeiser (1994)]{Dermer94} Dermer, C. D., \&
  Schlickeiser, R. 1994, ApJS, 90, 945

\bibitem[De Young (2003)]{Deyoung03} De Young, D. S. 2003, MNRAS, 343, 719

\bibitem[Fabian (1994)]{Fabian94} Fabian, A. C. 1994, ARA\&A, 32, 277

\bibitem[Fabian \etal\ (2000)]{Fabian00} Fabian, A. C., et al. 2000,
  MNRAS, 318, L65.

\bibitem[Fabian \etal\ (2002)]{Fabian02} Fabian, A. C., Celotti, A.,
 Blundell, K. M., Kassim, N. E., \&  Perley, R. A. 2002, MNRAS, 331, 369

\bibitem[Fabian \etal\ (2003)]{Fabian03} Fabian, A. C., Sanders,
 J. S., Allen, S. W., Crawford, C. S., Iwasawa, K., Johnstone, R. M.,
 Schmidt, R. W., \ Taylor, G. B. 2003, MNRAS, 344, L43

\bibitem[Felten \& Morrison (1966)]{Felten66} Felten, J. E. \&
  Morrison, P. 1966, ApJ, 146, 686

\bibitem[Flanagan \etal\ (2000)]{Flanagan00} Flanagan, K. A., Canizares,
  C. R., Davis, D. S., Dewey, D., Houck, J. C., Schattenburg,
  M. L. 2000, Bull. AAS, 32, 725

\bibitem[Fredericks \etal\ (2001)]{Fredericks01} Fredericks, A. C.,
  Flanagan, K. A., Canizares, C. R.,  Dewey, D., Houck, J. C., Schattenburg,
  M. L. 2001, Bull. AAS, 33, 1491


\bibitem[Gaetz \etal\ (2000)]{Gaetz00} Gaetz, T. J., Butt, Yousaf M.,
 Edgar, Richard J.,  Eriksen, Kristoffer A., Plucinsky, Paul P.,
 Schlegel, Eric M., Smith, Randall K. 2000, ApJ, 534, L47

\bibitem[Gaensler (2003)]{Gaensler03} Gaensler, B. M. 2003, in ``Texas in
      Tuscany (XXI Symposium on Relativistic Astrophysics),''
      eds. R. Bandiera, R. Maiolino, F. Mannucci, (World Scientific:
      Singapore),  297

\bibitem[Ghisellini \& Celotti (2001)]{Ghisellini01}  Ghisellini,
  G. \& Celotti, A. 2001, MNRAS, 327,  743

\bibitem[Gurvits \etal\ (2003)]{Gurvits03} Gurvits, Leonid I., Frey,
  Sandor, Mosoni, Laszlo, Garrington, Simon T., Garrett, Michael A.,
  \& Tsvetanov, Zlatan 2003, ``Maps of the Cosmos,'' IAU Symposium 216


\bibitem[Hansen \etal\ (1995)]{Hansen95} Hansen, L., J\o rgenson,
  H. E., \& N\o rgaard-Nielson, H. U. 1995, A\&A, 297, 13


\bibitem[Hardcastle \etal\ (2002)]{Hardcastle02} Hardcastle, M. J.,
  Birkinshaw, M., Cameron, R. A., Harris, D. E., Looney, L. W., \&
  Worrall, D. M. 2002, ApJ, 581, 948

\bibitem[Hardcastle \etal\ (2001)]{Hardcastle01} Hardcastle, M. J.,
  Birkinshaw, M. \& Worrall, D. M. 2001, MNRAS, 323, L17

\bibitem[Harris \etal\ (2000)]{Harris00} Harris, D. E. et al. 2000, ApJ, 530, L81

\bibitem[Heinz \etal\ (1998)]{Heinz98} Heinz, S., Reynolds, C. S., \&
  Begelman, M. C. 1998, ApJ, 501,126

\bibitem[Helfand \etal\ (2003)]{Helfand03} Helfand, D. J., Collins, B.
  F., \&  Gotthelf, E. V. 2003, ApJ, 582, 783

\bibitem[Hook \& McMahon (1998)]{Hook98} Hook, I. M., \& McMahon,
  R. G. 1998, MNRAS, 294, L7

\bibitem[Kellogg \etal\ (2001)]{Kellogg01} Kellogg, E., Pedelty, A. \&
  Lyon, R. G. 2001, ApJ, 563, L151

\bibitem[Marshall \etal\ (2001)]{Marshall01} Marshall, H. L. \etal\
  2001, ApJ, 549, L167

\bibitem[Marshall \etal\ (2002)]{Marshall02} Marshall, H. L. \etal\
  2002, Bull. AAS, 34, 647

\bibitem[McNamara (1995)]{Mcnamara95} McNamara, B. R., 1995, ApJ, 443, 77

\bibitem[McNamara \etal\ (2000)]{Mcnamara00} McNamara, B. R.,  \etal\
  2000, ApJ, 534, L135 

\bibitem[McNamara \etal\ (2001)]{Mcnamara01} McNamara, B. R., \etal\
  2001, ApJ, 562, L149

\bibitem[Meier (2003)]{Meier03} Meier, D. L. 2003, New Astronomy
  Reviews, 47, 667

\bibitem[Michel (1969a)]{Michel69a} Michel, F. C. 1969, ApJ, 157, 1183

\bibitem[Michel (1969b)]{Michel69b} Michel, F. C. 1969, ApJ, 158, 727

\bibitem[Nulsen \etal\ (2002)]{Nulsen02} Nulsen, P.E.J., David, L.P,
  McNamara, B. R., Jones, C., Forman, W. R., \& Wise, M. 2002, ApJ,
  568, 163

\bibitem[Nulsen (2003)]{Nulsen03} Nulsen, P. 2003, in "The Riddle of
  Cooling Flows in Galaxies and Clusters of Galaxies," eds
  T.H. Reiprich, J.C. Kempner, N. Soker,
  http://www.astro.virginia.edu/coolflow, and  astro-ph/0310195 

\bibitem[Pavlov \etal\ 2003]{Pavlov03} Pavlov,  G. G., Teter, M. A., Kargaltsev, O., \& Sanwal, D. 2003, ApJ, 591,1157

\bibitem[Sambruna \etal\ (2002)]{Sambruna02} Sambruna, R. \etal\ 2002,
  ApJ, 571, 206

\bibitem[Sanders \& Fabian (2002)]{Sanders02} Sanders, J. S., \&
  Fabian, A. C. 2002, MNRAS, 331, 273

\bibitem[Schwartz \etal\ (2000)]{Schwartz00} Schwartz, D. A. \etal\
  2000, ApJ, 540 L69

\bibitem[Schwartz (2002)]{Schwartz02} Schwartz, D. A. 2002, ApJ, 569, L23

\bibitem[Schwartz \etal\ (2003)]{Schwartz03} Schwartz, D. A. \etal\
  2003, in ``Galactic Nuclei: from Central Engine to Host Galaxy,''
  eds.: S. Collin, F. Combes \& I. Shlosman, ASP Conference Series,
  290, 359

\bibitem[Schwartz \etal\ (2003b)]{Schwartz03b} Schwartz, D. A. \etal\
  2003, New Astronomy Reviews, 47, 461

\bibitem[Siemiginowska \etal\ (2003)]{Siemiginowska03} Siemiginowska,
  A. \etal\ 2003, ApJ, 598, L15

\bibitem[Sun \etal\ (2003)]{Sun03} Sun, M., Jones, C., Murray, S. S.,
  Allen, S. W., Fabian, A. C., \& Edge, A. C. 2003, ApJ, 587, 619

\bibitem[Tavecchio \etal\ (2000)]{Tavecchio00} Tavecchio, F. \etal\
  2000, ApJ, 544, L23

\bibitem[Taylor \etal\ (1990)]{Taylor90} Taylor, G. B., Perley, R. A.,
  Inoue, M., Kato, T., Tabara, H., \& Aizu, K. 1990, ApJ, 360, 41

\bibitem[Tucker \& Rosner (1983)]{Tucker83} Tucker, W. H., \& Rosner,
  R. 1983, ApJ, 267, 547

\bibitem[Vignali \etal\ (2003)]{Vignali03} Vignali, C., Brandt, W. N.,
  Schneider, D. P., Garmire, G. P., \& Kaspi, S. 2003, AJ, 125, 418

\bibitem[Weisskopf \etal\ 2000]{Weisskopf00} Weisskopf, M. , \etal\
  2000, ApJ, 536, L81


\bibitem[White \etal\ (1997)]{White97} White, D. A., Jones, C., \&
  Forman, W. 1997, MNRAS, 292, 419

\bibitem[Wilson \etal\ (2000)]{Wilson00} Wilson, A.S., Young, A.J., \& Shopbell,P.L., 2000, Ap.J., 544, L27

\bibitem[Worrall \etal\ (2001)]{Worrall01} Worrall, D. M., Birkinshaw,
  M., Hardcastle, M. J. 2001, MNRAS, 326, L7


\end{thebibliography}
\end{document}